\documentclass[sigconf]{acmart}

\usepackage{svg}
\usepackage{mfirstuc}
\usepackage[utf8]{inputenc}
\usepackage{balance} 
\usepackage{afterpage}

\usepackage{makecell}
\usepackage{subcaption}
\usepackage{xcolor}
\usepackage{amssymb}  
\usepackage{arydshln}
\usepackage{multirow}
\usepackage{graphicx}
\usepackage{hyperref}
\newcommand{\etal}{\textit{et al.}}

\AtBeginDocument{%
  }

\copyrightyear{2025}
\acmYear{2025}
\setcopyright{acmlicensed}
\acmConference[MM '25] {Proceedings of the 33rd ACM International Conference on Multimedia}{October 27--31, 2025}{Dublin, Ireland.}
\acmBooktitle{Proceedings of the 33rd ACM International Conference on Multimedia (MM '25), October 27--31, 2025, Dublin, Ireland}
\acmISBN{979-8-4007-2035-2/2025/10}
\acmDOI{10.1145/3746027.3758191}
\begin{document}

%\title{GroundLie360: A Real-World Text-Video Misinformation Grounding Dataset}

% \title[A New Dataset and Benchmark for Grounding Multimodal Misinformation]
% {
% \raisebox{-0.6\height}{\includegraphics[height=3em]{figures/icon.png}} A New Dataset and Benchmark for Grounding \\ \vspace{-8mm} Multimodal Misinformation}
\title{A New Dataset and Benchmark for Grounding Multimodal Misinformation}

\author{Bingjian Yang}
\affiliation{%
  \institution{National Engineering Research Center for Multimedia Software, School of Computer Science, Wuhan University}
  \city{Wuhan}
  \country{China}
}

\author{Danni Xu}
% \email{dannixu@u.nus.edu}
\affiliation{%
  \institution{School of Computing, National University of Singapore}
  % \city{Singapore}
  \country{Singapore}
}

\author{Kaipeng Niu}
%\email{-}
\affiliation{%
  \institution{National Engineering Research Center for Multimedia Software, School of Computer Science, Wuhan University}
  \city{Wuhan}
  \country{China}
}

\author{Wenxuan Liu}
\affiliation{%
  \institution{School of Computer Science, Peking University; State Key Laboratory for Multimedia Information Processing, Peking University }
  \city{Beijing}
  \country{China}
}

\author{Zheng Wang}
\authornote{Corresponding author.}
\affiliation{%
  \institution{National Engineering Research Center for Multimedia Software, School of Computer Science, Wuhan University}
  \city{Wuhan}
  \country{China}
}

\author{Mohan Kankanhalli}
\affiliation{%
  \institution{School of Computing, National University of Singapore}
  % \city{Singapore}
  \country{Singapore}
}

\renewcommand{\shortauthors}{Bingjian Yang et al.}
\newcommand\datasetabbr{GroundLie360}
\newcommand\faketypeone{factual distortion}
\newcommand\Faketypeone{Factual distortion}
\newcommand\faketypetwo{spatial manipulation}
\newcommand\Faketypetwo{Spatial manipulation}
\newcommand\faketypethree{temporal manipulation}
\newcommand\Faketypethree{Temporal manipulation}
\newcommand\faketypefour{false speech}
\newcommand\Faketypefour{False speech}
\newcommand\faketypefive{contradictory content}
\newcommand\Faketypefive{Contradictory content}
\newcommand\faketypesix{unsupported content}
\newcommand\Faketypesix{Unsupported content}
\begin{abstract}
%The proliferation of online misinformation videos poses serious societal risks. While multiple misinformation video datasets and detection techniques have been proposed, they are devised for binary classification or single-modality localization based on post-processed data, lack of the interpretability needed to combat misinformation with high persuasion. In this paper, we introduce the task of Grounded Text-Video Fact-Checking, which aims not only to verify multimodal content but also to localize misleading segments across modalities. To support this task, we present the first real-world grounded misinformation dataset, \textsc{GroundLie360}, featuring: (1) a taxonomy of multimodal misinformation types, (2) fine-grained annotations of deception type and location across text, audio, and visual modalities, and (3) validation with evidence from Snopes and annotator reasoning. Moreover, we propose a novel VLM-based, QA-driven baseline, named FakeMark, that leverages both single-modality and cross-modality cues, enabling effective identification and grounding of misinformation across diverse modalities. Our experiments demonstrate the challenges of Grounded Text-Video Fact-Checking task. This work provides a foundation for explainable, interpretable multimodal misinformation detection.

The proliferation of online misinformation videos poses serious societal risks. Current datasets and detection methods primarily target binary classification or single-modality localization based on post-processed data, lacking the interpretability needed to counter persuasive misinformation. In this paper, we introduce the task of Grounding Multimodal Misinformation (GroundMM), which verifies multimodal content and localizes misleading segments across modalities. We present the first real-world dataset for this task, \textsc{GroundLie360}, featuring a taxonomy of misinformation types, fine-grained annotations across text, speech, and visuals, and validation with Snopes evidence and annotator reasoning. We also propose a VLM-based, QA-driven baseline, \textit{FakeMark}, using single and cross-modal cues for effective detection and grounding. Our experiments highlight the challenges of this task and lay a foundation for explainable multimodal misinformation detection. Dataset will be released at \href{https://github.com/yangbingjian/GroundLie360}{\textcolor{magenta}{https://github.com/yangbingjian/GroundLie360}}.

\end{abstract}

\vspace{-8mm}

%\ccsdesc[300]{Information systems~Social networks}

 %, Video Misinformation

\begin{teaserfigure}
  \includegraphics[width=\textwidth]{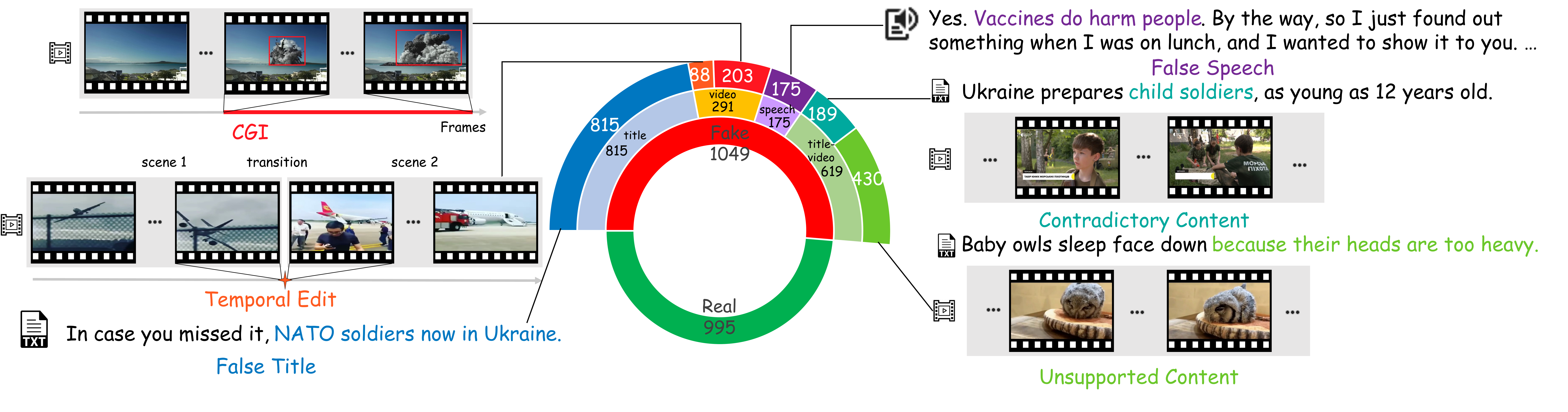}
\vspace{-8mm}
\caption{\textnormal{\textbf{Overview of the \textsc{GroundLie360} Dataset.} Our multi-modal benchmark contains 2,000+ fact-checked videos with fake type and grounding annotations. Fake types include: (1) {\textcolor[HTML]{2577C1}{False Title}/\textcolor[HTML]{752995}{False Speech}} - video title or spoken content containing demonstrably false claims; (2) {\textcolor[HTML]{FB6A36}{Temporal Edit}} - videos altered to distort event chronologies or fabricate deceptive narratives; (3) {\textcolor[HTML]{FB2320}{CGI}} - digitally manipulated or generated synthetic media; (4) {\textcolor[HTML]{1DA99E}{Contradictory Content}} - text-video semantic mismatches; and (5) {\textcolor[HTML]{6BC72B}{Unsupported Content}} - headlines lacking evidentiary support in video content. The dataset offers a unified benchmark for fake content classification and localization.}}
  % \Description{Enjoying the baseball game from the third-base
  % seats. Ichiro Suzuki preparing to bat.}
  \label{dataset_overview}
\end{teaserfigure}

% \begin{figure*}
%   \includegraphics[width=\textwidth]{acmart-primary/figures/Dataset_overview.pdf}
%   \caption{Overview of \textsc{GroundLie360} Dataset}
%   % \Description{Enjoying the baseball game from the third-base
%   % seats. Ichiro Suzuki preparing to bat.}
%   \label{fig:dataset_overview}
% \end{figure*}

\maketitle
\section{Introduction}

With the rise of social media, multimodal misinformation has become increasingly persuasive and deceptive, significantly influencing public opinion and behavior~\cite{vaccari2020deepfakes, westerlund2019emergence}.
Simple real/fake checks are no longer sufficient~\cite{blanchar2024trump}. Instead, detailed fact-checking fosters trust and enhances users' ability to recall and recognize misinformation even after extended periods~\cite{Porter2021}.
However, human-driven fact-checking remains labor-intensive and inefficient.

Consequently, the task of Detecting and Grounding Multi-Modal Media Manipulation ($\mathrm{DGM}^{4}$)~\cite{shao2023dgm4, shao2024dgm4++} was introduced, emphasizing the need to both detect the authenticity of multimodal media and ground manipulated content (\textit{i.e.,} image bounding boxes and text tokens). However, existing $\mathrm{DGM}^{4}$ efforts are limited to text-image pairs and a narrow range of misinformation types. In this work, we %define and construct a new task
propose a new and broader task: \textbf{Grounding Multimodal Misinformation (GroundMM)}. 
GroundMM extends the scope to visuals, speech, and text, requiring systems to not only detect misinformation but also ground and explain specific false elements across multiple modalities. 

The proposed \textbf{GroundMM} task brings three key challenges: 
1) Type confusion: Misinformation types often co-occur (\textit{e.g.}, miscaptioning with deepfake manipulation), making them hard to separate.
2) Multimodal complexity: Misinformation spans text, visuals, and speech, with complex dependencies that make grounding more difficult.
3) Annotation subjectivity: Real-world misinformation is subtle and context-sensitive, leading to annotation inconsistencies.

To address these challenges, we propose a \textbf{three-level annotation scheme} for constructing a GroundMM dataset. 
To resolve type confusion, each multimodal post is decomposed into its textual, visual, and cross-modal components, with modality-specific labels assigned accordingly. 
To handle misinformation complexity, annotators follow a hierarchical workflow: first determining the overall veracity, then assessing the veracity of each modality, and finally providing fine-grained grounding by marking the exact text span, frame indices, or bounding box where the falsehood appears.
To mitigate annotation subjectivity, all annotations are supported by verified fact-checking articles from \textit{Snopes}, and annotators provide justifications for each decision.

Based on the scheme, we present \textsc{GroundLie360} (Figure~\ref{dataset_overview}), the first unified \textbf{GroundMM} benchmark with the following characteristics:~\textbf{3-level annotation}: binary-type, six fake types and grounding;~\textbf{fine-grained multi-modal grounding}: involving video, speech, and title modalities, and covering token-level, temporal, and visual bounding-boxes grounding.

In addition, existing methods lack grounding for misinformation in complex multimodal video posts, particularly within real-world news contexts. To address this gap, we propose a new benchmark, \textit{FakeMark}, which integrates 
video processing tools and multimodal LLM in a question-driven pipeline. Through this design, we achieve fine-grained grounding across modalities, including token-level textual grounding, frame-index-level temporal grounding, and bounding-box-level visual grounding.  Key contributions of the paper include:
\begin{itemize}
\item \textbf{Task definition:} We introduce the novel task of Grounding Multimodal Misinformation, 
where models detect and localize misinformation across visual, auditory, and textual modalities.
\item \textbf{Dataset:} We create a new dataset with real-world news videos, featuring fine-grained annotations of misinformation types and grounding supported by expert fact-checking resources (\textit{e.g.}, Snopes).
\item \textbf{Methodology:} We develop a question-driven detection approach using VLM agents, enabling interpretable and localized detection of multimodal misinformation.
 
\end{itemize}
 
\section{Related Work}

\begin{table*}[!ht]
  \centering
    \caption{\textnormal{\textbf{Comparison of Multimodal Fake News Datasets.} Label Levels (L1: binary veracity classification. L2: fake types, L3: fake content grounding. M.A. (Manual Annotation), Annotation (Title: Textual headline, Speech: Transcript of verbal content, V.T.: Video Temporal localization, V.S.: Video Spatial localization). Range (RW-Gen: Real-World General Misinformation, DF: Deepfake, Tam: Tampering), Rat. (Explanation Rationale), and \#Plat. (Number of platforms from which the video data was collected).}}
  % SG-DF: Synthetically Generated Deepfake, SG-Tam: Synthetically Generated Tampering

\label{tab:dataset_comparison}
  \setlength{\tabcolsep}{3pt} 
  \renewcommand{\arraystretch}{1}
  \resizebox{\textwidth}{!}{
  \begin{tabular}{lcccccccccccccccc}
    \toprule
    \multirow{2}{*}{\textbf{Dataset}} & \multirow{2}{*}{\textbf{Year}} & \multirow{2}{*}{\textbf{Data Scale}} & \multicolumn{3}{c}{\textbf{Label Levels}} & \multirow{2}{*}{\textbf{M.A.}} & \multicolumn{4}{c}{\textbf{Annotation}} & \multicolumn{2}{c}{\textbf{Range}} & \multirow{2}{*}{\textbf{Rat.}} & \multirow{2}{*}{\textbf{\#Plat.}} \\
    \cmidrule(lr){4-6} \cmidrule(lr){8-11} \cmidrule(lr){12-13}
    & \textbf{(2000+)} & \textbf{(True/False)} & \textbf{L1} & \textbf{L2} & \textbf{L3} & & \textbf{Title} & \textbf{Speech} & \textbf{V.T.} & \textbf{V.S.} & \textbf{RW-Gen} & \textbf{DF}  & & \\
    \midrule
    \textsc{Official-NV}~\cite{wang2024official} & 21-24 & 5000T / 5000F & $\checkmark$ & 2 & $\times$ & $\times$ & $\times$ & $\times$ & $\times$ & $\times$ & $\times$ & $\times$  & $\times$ & 3 \\
    \textsc{FakeSV}~\cite{qi2023fakesv} & 17-22 & 1827T / 1827F & $\checkmark$ & 0 & $\times$ & $\times$ & $\times$ & $\times$ & $\times$ & $\times$ & $\checkmark$ & $\times$  & $\times$ & 2 \\
    \textsc{TRUE}~\cite{niu2025pioneering} & 16-24 & 1097T / 1828F & $\checkmark$ & 5 & $\times$ & $\times$ & $\checkmark$ & $\times$ & $\times$ & $\times$ & $\checkmark$ & $\times$  & $\checkmark$ & 8 \\
    \textsc{FakeNVE}~\cite{chen2025multimodalfakenewsvideo} & 17-22 & 1812T / 1802F & $\checkmark$ & 0 & $\times$ & $\checkmark$ & $\checkmark$ & $\times$ & $\times$ & $\times$ & $\checkmark$ & $\times$  & $\checkmark$ & 2 \\
    \textsc{DF-Platter}~\cite{Narayan_2023_CVPR} & 23 & 764T / 132,496DF & $\checkmark$ & 0 & $\checkmark$ & $\times$ & $\times$ & $\times$ & $\times$ & $\checkmark$ & $\times$ & $\checkmark$  & $\times$ & 1 \\
    \textsc{AV-Deepfake1M}~\cite{cai2024av} & 23 & 286,721T / 860,039DF & $\checkmark$ & 0 & $\checkmark$ & $\times$ & $\times$ & $\checkmark$ & $\checkmark$ & $\times$ & $\times$ & $\checkmark$  & $\times$ & 1 \\
    % MVTamperBench & 24 & 0T / 17,435Tam & $\times$ & 5 & $\checkmark$ & $\times$ & $\times$ & $\times$ & $\checkmark$ & $\times$ & $\times$ & $\times$ & $\checkmark$ & $\times$ & - \\
    \midrule
    \textsc{DGM}$^{4}$~\cite{shao2024dgm4++} (Image) & 24 & 77k+T/152k+Tam & $\checkmark$ & $\checkmark$ & $\checkmark$ &  $\times$ &  $\checkmark$ & $\times$ & $\times$ & $\times$ & $\times$ & $\checkmark$ & $\checkmark$ & 4 (human news)\\
    \midrule
    \textsc{GroundLie360} & 16-24 & 995T / 1049F & $\checkmark$ & \textbf{6} & $\checkmark$ & $\checkmark$ & $\checkmark$ & $\checkmark$ & $\checkmark$ & $\checkmark$ & $\checkmark$ & $\checkmark$ &  $\times$ & 8 \\
    \bottomrule
  \end{tabular}
  }

\end{table*}

\paragraph{Multimodal Misinformation Datasets} Research on video misinformation datasets has evolved from simple, coarse-grained labels to more explainable and fine-grained annotations. While existing multimodal misinformation datasets have laid essential groundwork, they still face key limitations. Early efforts such as \textsc{FakeSV}~\cite{qi2023fakesv} and \textsc{Official-NV}~\cite{wang2024official} offer only video-level binary labels. Subsequent datasets like \textsc{TRUE}~\cite{niu2025pioneering} and \textsc{FakeNVE}~\cite{chen2025multimodalfakenewsvideo} provide explanation-based annotations but lack precise grounding, offering only global justifications. More recent work, including \textsc{DF-Platter}~\cite{Narayan_2023_CVPR} and \textsc{AV-Deepfake1M}~\cite{cai2024av}, introduces spatial and temporal localization but focuses primarily on facial deepfakes, failing to capture the full spectrum of real-world misinformation. Shao~\etal~\cite{shao2023dgm4, shao2024dgm4++} introduced the Detecting and Grounding Multi-Modal Media Manipulation (DGM$^4$) task and dataset, targeting text-image manipulations focused on identity and emotional changes. However, DGM$^4$ is limited to synthetic, image-text content and narrow manipulation types (\textit{e.g.}, miscaptioning).
To address these gaps, we extend the task to the video-text domain and introduce a new dataset that uniquely integrates fine-grained textual, temporal and spatial grounding, covering a broad range of real-world fake types. This is the first comprehensive benchmark for grounding video misinformation, enhancing both interpretability and precision. Our dataset supports more credible and practical multimodal misinformation detection in real-world scenarios.
(See Table~\ref{tab:dataset_comparison} for a detailed comparison.)

\paragraph{Multimodal Misinformation Detection Methods} Approaches to video misinformation detection fall into two categories: traditional deep learning and LLM-based methods. Traditional models (\textit{e.g.}, VMID~\cite{zhong2024vmidmultimodalfusionLLM}, MultiTec~\cite{10854802}, MTPareto~\cite{yan2025mtparetomultimodaltargetedpareto}, MRGT~\cite{chen2025multimodalfakenewsvideo}) employ multimodal fusion, attention, and relational graphs for misinformation detection and explanation, but are limited by dataset constraints and black-box reasoning. LLM-based methods provide enhanced generalizability and interpretability. Evidence-based approaches like 3MFact~\cite{niu2025pioneering} and FIRE~\cite{xie-etal-2025-fire} combine retrieval and verification to fact-check claims. Fine-tuned models such as SNIFFER~\cite{Qi_2024_CVPR} enhance detection of out-of-context misinformation via tailored instruction tuning.
These trends signal a shift from task-specific architectures to generalizable LLM-based reasoning frameworks for tackling complex multimodal misinformation. However, when misinformation involves intricate visual, speech, and cross-modality cues, purely textual and global-level fact-checking fails to provide precise and comprehensive explanations. This work introduces a misinformation video grounding baseline combining vision-language models with prompt-based fact-checking and essential video tools.
% Their HAMMER++ model offers a transformer-based solution for detecting and grounding such edits.
% \subsection{Video Temporal Localization}

% \subsection{Deepfake Temporal Localization}

\section{GroundLie360 Dataset}

% We propose a new dataset for video misinformation, with the following key design goals:
% 1) Standardizing the task of misinformation localization by introducing a unified classification scheme that encompasses diverse types of misleading video content;
% 2) Defining grounding objectives and evaluation metrics for each category of video misinformation to support fine-grained, measurable benchmarking;
% 3) Implementing a three-level annotation framework that incorporates prevalent misinformation types through:
% Modality-specific labeling across text, visual, and cross-modal content; Joint annotation of misinformation type and its precise location (grounding); Reliability enhancement via fact verification from \textit{Snopes} and annotator-provided justifications.
 
Following previous fake news datasets~\cite{yao2023end,bu2024fakingrecipe}, the samples of Ground-Lie360 are derived from an authorized fact-checking website -- Snopes\footnote{\url{https://www.snopes.com/}}.
It covers 1,466 influential video-related events that have been systematically investigated. 
% This dataset features a three-level annotation scheme, including:  
% 1) binary veracity labels,  
% 2) six fake type labels, and  
% 3) fake content grounding annotations (see \textcolor{black}{Figure}~\ref{dataset_overview} for an overview of the dataset).

\subsection{Dataset Construction Process}
\begin{figure}[h]
  \centering
  \includegraphics[width=\linewidth]{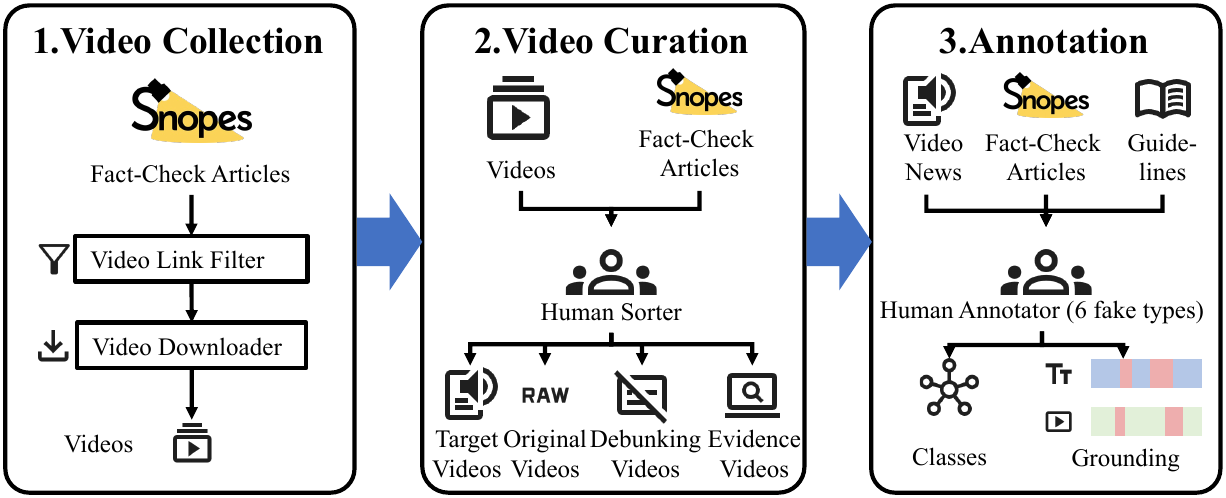}
\caption{\textnormal{\textbf{Construction pipeline of \textsc{GroundLie360}.} It consists of three stages: 
(1) \textbf{Video Collection} harvesting videos from snopes.com; (2) \textbf{Curation} separating videos to be annotated from auxiliary videos; (3) \textbf{Annotation} labeling target videos with 6 fake types and grounding information.}}
  \Description{}
  \label{fig:pipeline}
\end{figure}

The construction pipeline of \textsc{GroundLie360} is illustrated in Figure~\ref{fig:pipeline}.
For the annotation step, given the complexity of video misinformation, where key errors may occur in any modality or arise from cross-modal inconsistencies, we propose a three-level hierarchical annotation framework, as detailed in Table~\ref{tab:dataset_comparison}. Level 1 focuses on \textbf{binary veracity classification}, where each video is automatically labeled as Real or Fake based on fact-check ratings (\textit{e.g.}, from Snopes). 
Level 2 refers to the~\textbf{Fake Types}. Annotators identify six fake types (Figure~\ref{dataset_overview}) across four modalities. Specifically, the title modality checks whether the video headline is \textit{false title}, and speech examines if the spoken content includes \textit{false speech}. The video modality assesses the presence of \textit{temporal edit} or \textit{CGI}, while cross-modal inconsistencies evaluate whether the video-text pair exhibits \textit{contradictory} or \textit{unsupported} content.
In Level 2, each video may be assigned multiple labels within or across modalities, enabling fine-grained analysis of fake content and enhancing dataset reusability for downstream tasks. Level 3,~\textbf{Fake Content Grounding} aims to ground the misinformation source for each detected fake type. A summary of grounding contents is shown in Table~\ref{tab:grounding-types}, with more details provided in the supplementary material at the dataset link.

\begin{table}[h]
\footnotesize
\centering
\caption{\textnormal{\textbf{Grounding Content for Each Fake Type.} Text: Textual Grounding, V.T.: Video Temporal Grounding, V.S.: Video Spatial Grounding}}
\vspace{-2mm}
  \resizebox{\columnwidth}{!}{
\begin{tabular}{lll}
\toprule
\textbf{Fake Modality} & \textbf{Fake Type} & \textbf{Grounding Modality and Content} \\
\midrule
Text (Title) & False Title & \textbf{Text}: False text span \\
\midrule
Speech & False Speech & \textbf{Speech}: False text span \\
\midrule
\multirow{2}{*}{Text-Video} & Contradictory Content & \textbf{Text}: Contradictory text \\
           & Unsupported Content & \textbf{Text}: Unsupported text \\
\midrule
\multirow{2}{*}{Video} & Temporal Edit & \textbf{V.T.}: Edited timestamps \\
      & CGI & \textbf{V.T.}: CGI time span; \textbf{V.S.}: CGI region \\
\bottomrule
\end{tabular}}
\label{tab:grounding-types}
\vspace{-2mm}
\end{table}

\subsection{Quality Control}
To reduce individual bias, each sample is independently annotated by two annotators. Disagreements on binary veracity and fake types are resolved by a third expert who assigns the final label. We evaluate inter-annotator agreement (IoU) on grounding, achieving 0.74 for text, 0.84 for temporal spans, and 0.98 for bounding boxes. All scores exceed 70\%, indicating high data quality.

\subsection{Statistical Analysis}
\begin{figure*}[t]
  \centering
  \includegraphics[width=\linewidth]{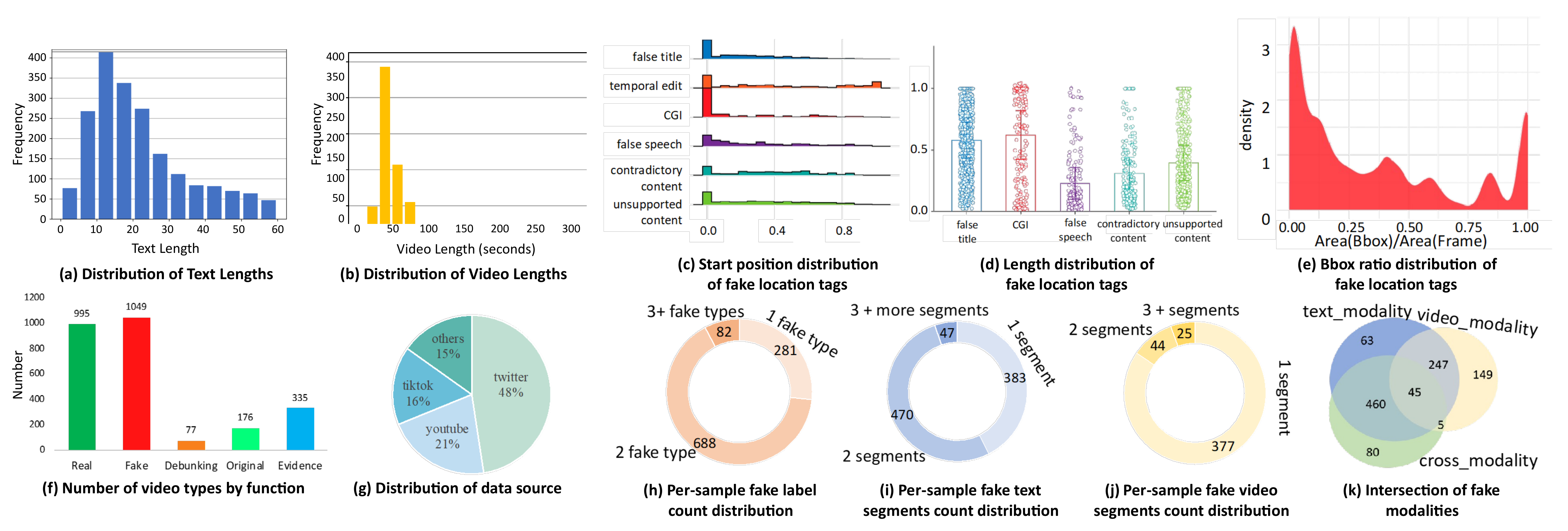}
  \caption{\textnormal{\textbf{Statistic of \textsc{GroundLie360}.} (d) excludes the temporal edit type from analysis, as isolated timestamps cannot be incorporated into proportion-based duration calculations.}}
  \label{fig:statistic}
  \vspace{-4mm}
\end{figure*}

\textsc{GroundLie360} consists of 2,044 target videos under 5 minutes and 588 auxiliary videos (original, debunking, and evidence videos). Figure~\ref{fig:statistic}(a), (b), (f), and (g) present basic dataset statistics, including text/video length distribution and source/video type distribution.

We conduct a series of statistical analyses on dataset to provide insights into multimodal video misinformation detection. As shown in Figure~\ref{dataset_overview}~\footnote{Fake type counts reflect the number of videos labeled with each type; totals may exceed the number of videos due to multi-labeling. Modality counts are the sum of fake types within each modality. For instance, \#V-T = \#Contradictory + \#Unsupported.}, the quantities across different modalities are relatively balanced. 
Figure~\ref{fig:statistic}(c), (d), (e) illustrate the distribution of grounding locations for the annotated fake segments:
In Figure~\ref{fig:statistic}(c), most fake segments start near the beginning of the video. However, for the temporal edit type, fake segments are more evenly distributed across the timeline, with manipulation at the end also being common — suggesting that \textit{truncated clip} is a frequent strategy.
As shown in Figure~\ref{fig:statistic}(d), fake durations are generally less than half the total video length. This is especially true for false speech, where short snippets often suffice to convey a misleading narrative or claim.
Figure~\ref{fig:statistic}(e) shows that most grounding bounding boxes (BBoxes) occupy a very small portion of the frame (typically < 0.25). Several distinct peaks can be observed, potentially corresponding to common manipulation targets such as humans or faces.
Additionally, there is a non-negligible number of cases where the entire frame is manipulated, indicating that fully AI-generated or synthetically altered videos are also prevalent in real-world scenarios.

Figure~\ref{fig:statistic}(h)-(k) illustrate the distribution of the number of fake types and fake segments per sample. The majority of samples (>60\%) contain two fake types (Figure~\ref{fig:statistic}(h)), and when combined with Figure~\ref{fig:statistic}(i), it suggests that cross-modality inconsistencies frequently co-occur with false textual content.
A non-negligible portion of samples (>7\%) contain three or more fake types, indicating the presence of more complex or multi-faceted manipulations.
The number of fake segments in text is primarily concentrated around one or two per sample (Figure~\ref{fig:statistic}(j)), whereas in video (Figure~\ref{fig:statistic}(k)), manipulations are limited to a single segment (>80\%), highlighting that single-segment tampering is the dominant pattern.

\begin{figure*}[htbp]
    \centering
    \includegraphics[width=\textwidth]{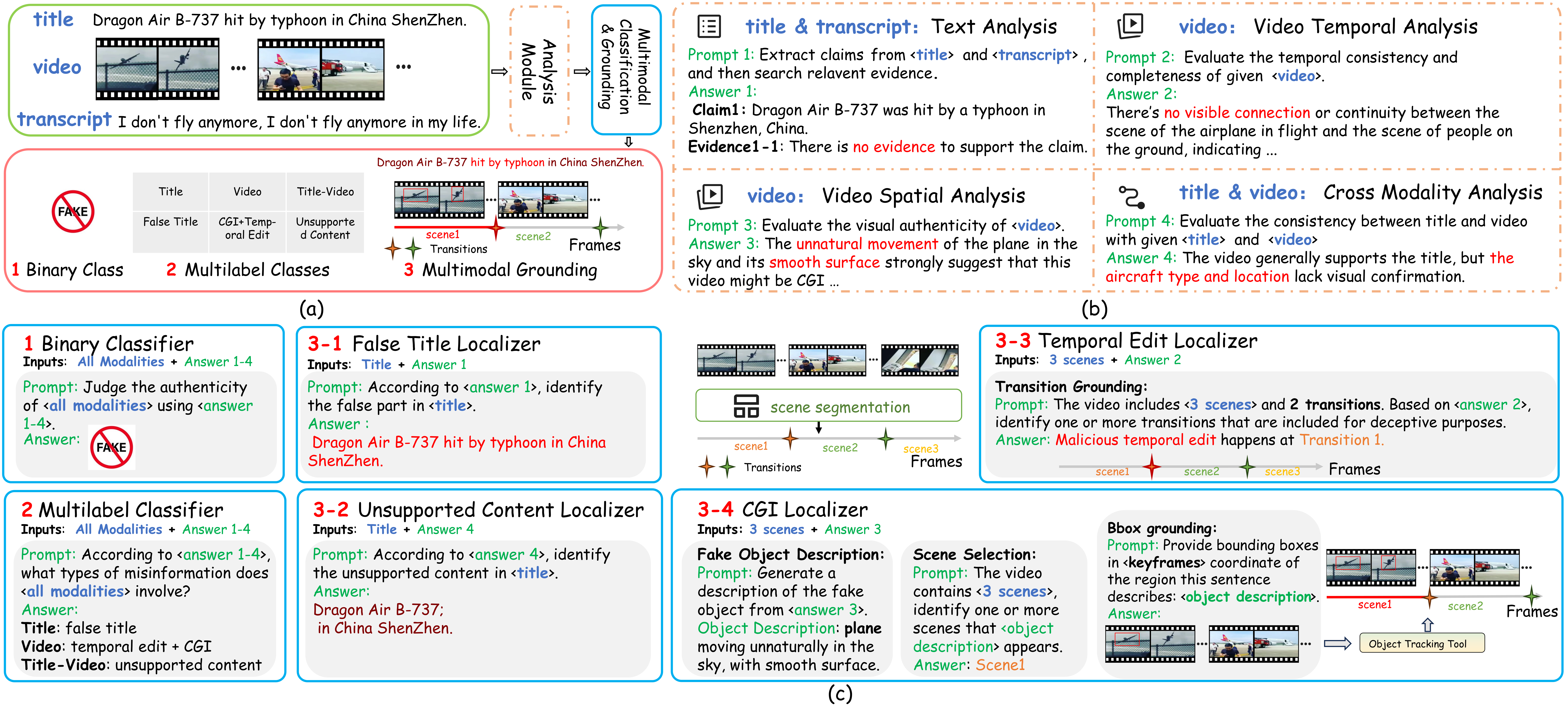}
    \Description{Diagram of the proposed framework for multimodal video fake news detection and grounding.}
    \captionsetup{font=small}
    \vspace{-8mm}
    \caption{\textnormal{\textbf{The workflow of \textit{FakeMark}.} (a) Overall pipeline; (b) Analysis Module, including text, video temporal/spatial, and cross-modality analyses; (c) Multimodal Classification and Grounding with a binary classifier, a multilabel classifier, and four localizers (false speech and contradictory content share designs with false title and unsupported types, respectively).}}
    \label{fig:framework}
\vspace{-5mm}
\end{figure*}

\section{Method}

We propose \textit{FakeMark}: A unified LLM-VLM framework that, given a video with its title and transcript, simultaneously predicts veracity labels, identifies misinformation types, and localizes fake content across modalities (Figure~\ref{fig:framework}).
\paragraph{Problem Definition} Each instance $(x, y, \mathbf{c}, \mathbf{G})$ contains: (1) multimodal input $x=(\mathcal{V}, \mathcal{T}, \mathcal{R})$ (video, title, ASR transcript); (2) veracity label $y\in\{0,1\}$; (3) fake type vector $\mathbf{c}\in\{0,1\}^6$ (when $y=1$); (4) grounding annotations $\mathbf\{G\}=\{G_i|c_i=1\}$. The task requires: (i) binary veracity classification; (ii) multi-label fake type prediction for fake samples; (iii) grounding localization prediction for each detected fake type.

\subsection{Analysis Module}
The analysis module transforms each input sample $x$ into a set of textual analyses $\mathcal{A}$, which are used in downstream tasks. We design four analytical prompts to address the six defined types of misinformation as shown in Figure \ref{fig:framework}(b).~\textit{Text analysis}, targeting \textit{false title} and \textit{false speech}, takes the title ($\mathcal{T}$) and transcript ($\mathcal{R}$) as input: $\mathbf{A_t} = LLM(\mathcal{T}, \mathcal{R})$.~\textit{Video temporal analysis}, addressing \textit{temporal edits}, uses the video input ($\mathcal{V}$):
    $\mathbf{A_{vt}} = VLM(\mathcal{V})$.
\textit{Video spatial analysis}, designed for detecting \textit{CGI}, also operates on video input: $\mathbf{A_{vs}} = VLM(\mathcal{V})$.
\textit{Cross-modal analysis}, for identifying \textit{contradictory content} and \textit{unsupported content}, takes both the title and video as input: $\mathbf{A_c} = VLM(\mathcal{T}, \mathcal{V})$. 
The final analysis output is represented as: $\mathcal{A} = VLM(\mathcal{V}, \mathcal{T}, \mathcal{R})$.

\subsection{Multimodal Classification and Grounding}

\paragraph{Binary Classifier and Multilabel Classifier.} As illustrated in Figure~\ref{fig:framework}(c1) and (c2). Given the input sample $x$ and its corresponding analytical results $\mathcal{A}$, the binary authenticity prediction is formulated as $\hat{y} = VLM(x, \mathcal{A})$, while the multilabel classification for misinformation types is expressed as $\hat{\mathbf{c}} = VLM(x, \mathcal{A})$.

\paragraph{Textual Localizer.} It is applied to four misinformation types: \textit{false title}, \textit{false speech}, \textit{unsupported content}, and \textit{contradictory content}.

The grounding module takes as input the relevant textual modality (i.e., $\mathcal{T}$ or $\mathcal{R}$) along with the corresponding analysis result (i.e., $\mathbf{A_t}$ or $\mathbf{A_c}$), and outputs the grounded token spans $\hat{\mathcal{T}}$ as illustrated in Figure~\ref{fig:framework}(c3-1) and Figure~\ref{fig:framework}(c3-2). 

\paragraph{Temporal Edit and CGI localizer.} We first perform video scene segmentation using TransNetV2~\cite{soucek2024transnet} for two main reasons. First, for the \textit{temporal edit} type, the candidate timestamps for grounding align with scene transition points. Second, scene segmentation helps refine the grounding granularity for the \textit{CGI} type.
The scene transition generation process is defined as:
$
\Theta^t = \text{seg}(\mathcal{V}), \quad \Theta^t \subseteq \{1, 2, \dots, L\}
$, where $\Theta^t$ is the set of frame indices marking scene transitions, $L$ is the total number of video frames.
The temporal grounding process of \textit{temporal edit} is using VLM to select misleading transitions $\hat{\Theta}$, as shown in Figure~\ref{fig:framework}(c3–3).
The CGI-type misinformation requires both temporal and spatial grounding, thus the grounding output is defined as $\hat{G} = (\hat{\Theta}, \hat{B})$, representing the set of fake frames and bounding boxes. As shown in Figure~\ref{fig:framework}(c3--4), we first generate an object-level grounding query $q = \text{LLM}(\mathbf{A}_{vs})$. Given $q$ and scene transition frames $\Theta^t$, the VLM selects fake scene intervals $(f_s, f_e) = \text{LLM}(\mathcal{V}, q, \Theta^t)$, where $f_s, f_e \in \Theta^t$ and $f_s \leq f_e$. We uniformly sample representative frames as $\Theta^* = \text{UniformSample}(f_s, f_e)$. These frames are then used for initial spatial grounding: $B^* = \{ b_f \mid b_f \in \mathbb{R}^4,\; f \in \Theta^* \} = \text{LLM}(\mathcal{V}, q, \Theta^*)$. Finally, an object tracking model~\cite{ravi2024sam} refines both the bounding boxes and their temporal span, producing the final output: $\hat{G} = (\hat{\Theta}, \hat{B}) = \text{Track}(\Theta^*, B^*)$.

\section{Experiments}
The experiments aim to answer the following research questions: 
\textbf{Q1}: How well do recent methods perform on classification in \textsc{GroundLie360}? 
\textbf{Q2}: How effective is \textit{FakeMark} for multimodal grounding? 
\textbf{Q3}: How challenging is the GroundMM task across different fake types?

\subsection{Experimental settings}
\paragraph{Dataset Split.} We evaluate zero-shot methods on the entire Ground-Lie360 dataset. For supervised methods, the dataset is divided into 70\% for training, 15\% for validation, and 15\% for testing. 
\paragraph{Evaluation Metrics.} We use different evaluation metrics depending on the task type: 
(1) \textbf{Binary veracity classification:} Precision, Recall, and F1 score. 
(2) \textbf{Sub-type classification:} Macro-averaged Precision, Recall, and F1 score. 
(3) \textbf{Grounding tasks:} 
\textit{(a) Textual grounding:} Token-level Precision, Recall, and F1 score; 
\textit{(b) Video temporal grounding:} Frame-level Precision, Recall, and F1 score; 
\textit{(c) Video spatial-temporal grounding:} Mean temporal IoU (m\_tIoU), mean visual IoU (m\_vIoU), and vIoU@0.3 / vIoU@0.5~\cite{zhang2020does}.

\paragraph{Implementation Details.} 

We utilize GPT-4o-mini~\cite{achiam2023gpt} as the LLM for text-only tasks, and InternVL-8B~\cite{chen2024internvl} as the VLM for visual and multimodal tasks. The LLM accepts up to 1024 tokens as input. For VLM video inputs, both the entire video and individual scenes obtained via scene segmentation-are uniformly sampled to 16 frames. We use TransNetV2~\cite{soucek2024transnet} for scene segmentation and SAM2~\cite{ravi2024sam} for object tracking.

\subsection{Classification (Q1)}

\textit{Binary classification.} Table~\ref{tab:binary_results} shows binary classification performance, comparing our \textit{FakeMark} with SVFEND~\cite{fakesv} and FakingRecipe \cite{bu2024fakingrecipe}, both of which model cross-modal correlations to enhance news content representation. While \textit{FakeMark} may not outperform task-specific methods, it provides a unified, training-free framework that jointly handles multiple tasks within a single model.

\begin{table}[h]
\footnotesize
\centering
\captionsetup{font=small}
\caption{\textnormal{\textbf{Binary classification results.}~\textbf{Trad.} and~\textbf{LLM.} represent for traditional video fake detection methods and LLM-based method.}}
\begin{tabular}{lccccc}
\toprule
\textbf{Approach} & \textbf{Method} & \textbf{Acc.}  & \textbf{Prec.} & \textbf{Recall} & \textbf{F1} \\
\midrule
\multirow{2}{*}{\textbf{Trad.}} 
  & SVFEND         & 71.50 & 69.27 & 71.50 & 67.54 \\
  & FakingRecipe   & 76.14 & 75.06 & 74.84 & 74.94 \\
\midrule
\textbf{LLM.} & \textit{FakeMark}  & 62.90 & 67.81 & 52.81 & 59.38 \\
\bottomrule
\end{tabular}
\label{tab:binary_results}
\end{table}

\textit{Sub-type classification.} 
Table~\ref{tab:classification_text_grounding} presents classification performance across fake types. \textit{FakeMark} shows substantial variability across categories. It performs better on false title detection (62.40), likely due to its reliance on textual cues, but struggles with temporal edits (6.24), which require capturing subtle visual changes. Although both false speech and false title are text-based tasks, their performance diverges, likely due to sample imbalance and ambiguity in speaker identities.
For CGI-type content, high recall suggests strong sensitivity to spatiotemporal patterns, but low precision indicates over-reliance on coarse visual features.
Detecting contradictory and unsupported content involves cross-modal reasoning. The model performs poorly in both, reflecting challenges in aligning video with textual information, highlighting the difficulty of the \textsc{GroundLie360} benchmark.

\begin{table}[h]
\centering
\footnotesize
\captionsetup{font=small}
\caption{\textnormal{\textbf{Subtype classification and text grounding performance across fake types.}~\textbf{Prec.} stands for precision.}}
\vspace{-3mm}
 \resizebox{\columnwidth}{!}{
\begin{tabular}{lcccccc}
\toprule
\multirow{2}{*}{\textbf{Type}} & \multicolumn{3}{c}{\textbf{Classification}} & \multicolumn{3}{c}{\textbf{Text Grounding}} \\
\cline{2-7}
& \textbf{Prec.} & \textbf{Recall} & \textbf{F1} & \textbf{Prec.} & \textbf{Recall} & \textbf{F1} \\
\midrule
False title         & 62.40 & 46.01 & 52.97 & 49.45 & 33.06 & 39.63 \\
Temporal edit       & 6.24  & 32.95 & 10.49 & --    & --    & --    \\
CGI                 & 23.69 & 55.67 & 33.24 & --    & --    & --    \\
False speech        & 22.41 & 44.57 & 29.83 & 12.48 & 11.48 & 11.96 \\
Contradictory content       & 12.16 & 23.81 & 16.10 & 6.43  & 17.08 & 8.93  \\
Unsupported content & 31.11 & 34.65 & 32.78 & 20.20 & 16.19 & 17.97 \\
\hline
Micro average       & 28.80 & 41.53 & 34.01 & 23.91 & 21.05 & 22.39 \\
Macro average       & 26.33 & 39.61 & 29.23 & 22.14 & 19.45 & 19.62 \\
\hline
\end{tabular}}
\label{tab:classification_text_grounding}
\end{table}
\vspace{-4mm}

\subsection{Grounding (Q2)}

We plot the grounding performance of each fake type in Tables~\ref{tab:classification_text_grounding} and \ref{tab:video_grounding}. The results
deliver more interpretation as temporal-based detection is a challenge when grounding.

\paragraph{Textual grounding.}

Table~\ref{tab:classification_text_grounding} shows substantial variation in grounding performance across types. \textit{False title} achieves the highest F1 (39.63), benefiting from short, structured text. In contrast, \textit{false speech} performs poorly (11.96) due to ambiguous and fragmented spoken language. \textit{Contradictory} and \textit{unsupported} types, which require cross-modal reasoning, also show low F1 scores (8.93 and 17.97). Overall performance (F1 = 22.39) reflects the challenge of grounding misinformation in noisy or cross-modal text.

\paragraph{Video temporal grounding.}

The grounding schema includes two types with temporal grounding: \textit{temporal edit} and \textit{CGI}. While both use frames, the former grounds over scene transitions, and the latter considers all frames.
As shown in Table~\ref{tab:video_grounding}, both types yield low F1 scores (2.37 and 17.15), likely due to the multi-stage pipeline where grounding samples are narrowed down by binary and multi-class classification, and the insufficient information for distinguishing misleading temporal edits from normal ones.

\begin{table}[!htbp]
\footnotesize
\centering
\captionsetup{font=small}
\caption{\textnormal{\textbf{Performance on video grounding tasks.} Temporal grounding is evaluated with Precision (Prec.), Recall (Rec.), and F1; spatial-temporal grounding with m\_tIoU, m\_vIoU, and vIoU@ thresholds.~\textbf{T.E.} is temporal edit.}}
\vspace{-3mm}
\resizebox{\columnwidth}{!}{
\begin{tabular}{l@{\hskip 6pt}ccc@{\hskip 10pt}cccc}
\toprule
\textbf{Type} & \textbf{Prec.} & \textbf{Rec.} & \textbf{F1} & \textbf{m\_tIoU} & \textbf{m\_vIoU} & \textbf{vIoU@0.3} & \textbf{vIoU@0.5} \\
\hline
T. E. & 1.39 & 8.26 & 2.37 & -- & -- & -- & -- \\
CGI           & 12.98 & 25.26 & 17.15 & 13.05 & 7.74 & 9.41 & 7.36 \\
\bottomrule
\end{tabular}}
\label{tab:video_grounding}
\end{table}
\vspace{-3mm}
\paragraph{Video spatial temporal grounding.} 
The spatial grounding performance for the CGI type, as shown in Table~\ref{tab:video_grounding}, remains relatively low on the \textsc{GroundLie360} dataset. This suggests that direct prompting for CGI localization is still immature and insufficient for accurately identifying fake visual content. There remains significant room to explore how VLMs can better support spatial grounding of manipulated visuals.

\subsection{Case  Study (Q3)}
\begin{figure}[H]
  \centering
  \includegraphics[width=\linewidth]{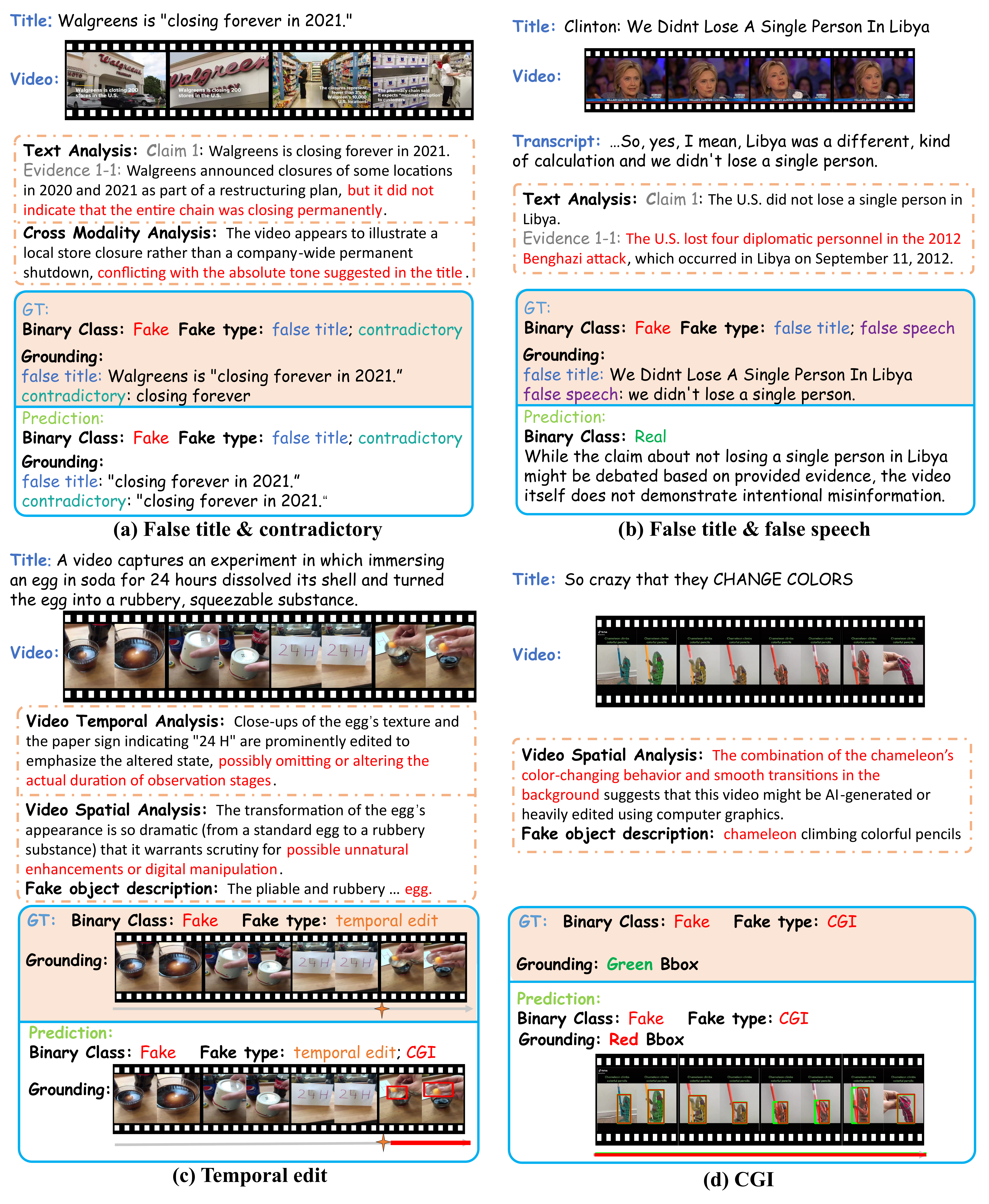}
  \vspace{-6mm}
  \caption{\textnormal{\textbf{Four representative cases in \textsc{GroundLie360}.} (a)(d) are successful cases about classification and grounding. (b) and (c) illustrate failure cases: (b) involves a binary misclassification, while (c) involves a multiclass misclassification.}}
  \Description{}
  \label{fig:casestudy}
  \vspace{-4mm}
\end{figure}

Our qualitative analysis reveals GroundMM’s core challenges. In Figure~\ref{fig:casestudy}(a) and (d), the model successfully performed binary classification and multi-label classification , achieving acceptable grounding despite imperfect token-level and bbox alignment. Figure~\ref{fig:casestudy}(b) and (c) respectively illustrate how errors in binary and multi-label classification stage are propagated downstream and directly magnify failures in the localization. These results highlight two key difficulties: (1) grounding requires deep understanding of varied misinformation patterns, and (2) its multi-stage nature ties grounding accuracy to upstream classification, making error propagation a critical issue.
\vspace{-4mm}

\section{Conclusion}

This work introduces a new task—Grounding Multimodal Misinformation (GroundMM), that advances the field from coarse-grained detection to fine-grained, modality-specific localization of misinformation. The proposed \textsc{GroundLie360} dataset offers the first comprehensive benchmark for this task, supporting research on multimodal explainable misinformation detection. By integrating video-language models into a unified pipeline, this work also demonstrates the potential and challenges of general-purpose models for complex, real-world misinformation challenges. The dataset and task definition are expected to benefit a wide range of applications, including fact-checking, general deepfake grounding, and cross-modal content analysis.

\bibliographystyle{ACM-Reference-Format}
\balance
\bibliography{main}


\begin{thebibliography}{25}

%%% the Web address.

\ifx \showCODEN    \undefined \def \showCODEN     #1{\unskip}     \fi
\ifx \showISBNx    \undefined \def \showISBNx     #1{\unskip}     \fi
\ifx \showISBNxiii \undefined \def \showISBNxiii  #1{\unskip}     \fi
\ifx \showISSN     \undefined \def \showISSN      #1{\unskip}     \fi
\ifx \showLCCN     \undefined \def \showLCCN      #1{\unskip}     \fi
\ifx \shownote     \undefined \def \shownote      #1{#1}          \fi
\ifx \showarticletitle \undefined \def \showarticletitle #1{#1}   \fi
\ifx \showURL      \undefined \def \showURL       {\relax}        \fi
% invisible to TeX
\providecommand\bibfield[2]{#2}
\providecommand\bibinfo[2]{#2}
\providecommand\natexlab[1]{#1}
\providecommand\showeprint[2][]{arXiv:#2}

\bibitem[Achiam et~al\mbox{.}(2023)]%
        {achiam2023gpt}
\bibfield{author}{\bibinfo{person}{Josh Achiam}, \bibinfo{person}{Steven Adler}, \bibinfo{person}{Sandhini Agarwal}, \bibinfo{person}{Lama Ahmad}, \bibinfo{person}{Ilge Akkaya}, \bibinfo{person}{Florencia~Leoni Aleman}, \bibinfo{person}{Diogo Almeida}, \bibinfo{person}{Janko Altenschmidt}, \bibinfo{person}{Sam Altman}, \bibinfo{person}{Shyamal Anadkat}, {et~al\mbox{.}}} \bibinfo{year}{2023}\natexlab{}.
\newblock \showarticletitle{Gpt-4 technical report}.
\newblock \bibinfo{journal}{\emph{arXiv preprint arXiv:2303.08774}} (\bibinfo{year}{2023}).
\newblock


\bibitem[Blanchar and Norris(2024)]%
        {blanchar2024trump}
\bibfield{author}{\bibinfo{person}{John~C. Blanchar} {and} \bibinfo{person}{Catherine~J. Norris}.} \bibinfo{year}{2024}\natexlab{}.
\newblock \showarticletitle{Trump, Twitter, and Truth Judgments: The Effects of “Disputed” Tags and Political Knowledge on the Judged Truthfulness of Election Misinformation}.
\newblock \bibinfo{journal}{\emph{HKS Misinformation Review}} (\bibinfo{date}{September} \bibinfo{year}{2024}).
\newblock
\urldef\tempurl%
\url{https://misinforeview.hks.harvard.edu/article/trump-twitter-and-truth-judgments-the-effects-of-disputed-tags-and-political-knowledge-on-the-judged-truthfulness-of-election-misinformation/}
\showURL{%
\tempurl}


\bibitem[Bu et~al\mbox{.}(2024)]%
        {bu2024fakingrecipe}
\bibfield{author}{\bibinfo{person}{Yuyan Bu}, \bibinfo{person}{Qiang Sheng}, \bibinfo{person}{Juan Cao}, \bibinfo{person}{Peng Qi}, \bibinfo{person}{Danding Wang}, {and} \bibinfo{person}{Jintao Li}.} \bibinfo{year}{2024}\natexlab{}.
\newblock \showarticletitle{Fakingrecipe: Detecting fake news on short video platforms from the perspective of creative process}. In \bibinfo{booktitle}{\emph{Proceedings of the 32nd ACM International Conference on Multimedia}}. \bibinfo{pages}{1351--1360}.
\newblock


\bibitem[Cai et~al\mbox{.}(2024)]%
        {cai2024av}
\bibfield{author}{\bibinfo{person}{Zhixi Cai}, \bibinfo{person}{Shreya Ghosh}, \bibinfo{person}{Aman~Pankaj Adatia}, \bibinfo{person}{Munawar Hayat}, \bibinfo{person}{Abhinav Dhall}, \bibinfo{person}{Tom Gedeon}, {and} \bibinfo{person}{Kalin Stefanov}.} \bibinfo{year}{2024}\natexlab{}.
\newblock \showarticletitle{AV-Deepfake1M: A large-scale LLM-driven audio-visual deepfake dataset}. In \bibinfo{booktitle}{\emph{Proceedings of the 32nd ACM International Conference on Multimedia}}. \bibinfo{pages}{7414--7423}.
\newblock
\href{https://doi.org/10.1145/3664647.3680795}{doi:\nolinkurl{10.1145/3664647.3680795}}


\bibitem[Chen et~al\mbox{.}(2025)]%
        {chen2025multimodalfakenewsvideo}
\bibfield{author}{\bibinfo{person}{Lizhi Chen}, \bibinfo{person}{Zhong Qian}, \bibinfo{person}{Peifeng Li}, {and} \bibinfo{person}{Qiaoming Zhu}.} \bibinfo{year}{2025}\natexlab{}.
\newblock \bibinfo{title}{Multimodal Fake News Video Explanation: Dataset, Analysis and Evaluation}.
\newblock
\showeprint[arxiv]{2501.08514}~[cs.CV]
\urldef\tempurl%
\url{https://arxiv.org/abs/2501.08514}
\showURL{%
\tempurl}


\bibitem[Chen et~al\mbox{.}(2024)]%
        {chen2024internvl}
\bibfield{author}{\bibinfo{person}{Zhe Chen}, \bibinfo{person}{Jiannan Wu}, \bibinfo{person}{Wenhai Wang}, \bibinfo{person}{Weijie Su}, \bibinfo{person}{Guo Chen}, \bibinfo{person}{Sen Xing}, \bibinfo{person}{Muyan Zhong}, \bibinfo{person}{Qinglong Zhang}, \bibinfo{person}{Xizhou Zhu}, \bibinfo{person}{Lewei Lu}, {et~al\mbox{.}}} \bibinfo{year}{2024}\natexlab{}.
\newblock \showarticletitle{Internvl: Scaling up vision foundation models and aligning for generic visual-linguistic tasks}. In \bibinfo{booktitle}{\emph{Proceedings of the IEEE/CVF conference on computer vision and pattern recognition}}. \bibinfo{pages}{24185--24198}.
\newblock


\bibitem[Narayan et~al\mbox{.}(2023)]%
        {Narayan_2023_CVPR}
\bibfield{author}{\bibinfo{person}{Kartik Narayan}, \bibinfo{person}{Harsh Agarwal}, \bibinfo{person}{Kartik Thakral}, \bibinfo{person}{Surbhi Mittal}, \bibinfo{person}{Mayank Vatsa}, {and} \bibinfo{person}{Richa Singh}.} \bibinfo{year}{2023}\natexlab{}.
\newblock \showarticletitle{DF-Platter: Multi-Face Heterogeneous Deepfake Dataset}. In \bibinfo{booktitle}{\emph{Proceedings of the IEEE/CVF Conference on Computer Vision and Pattern Recognition (CVPR)}}. \bibinfo{pages}{9739--9748}.
\newblock


\bibitem[Niu et~al\mbox{.}(2025)]%
        {niu2025pioneering}
\bibfield{author}{\bibinfo{person}{Kaipeng Niu}, \bibinfo{person}{Danni Xu}, \bibinfo{person}{Bingjian Yang}, \bibinfo{person}{Wenxuan Liu}, {and} \bibinfo{person}{Zheng Wang}.} \bibinfo{year}{2025}\natexlab{}.
\newblock \showarticletitle{Pioneering Explainable Video Fact-Checking with a New Dataset and Multi-role Multimodal Model Approach}. In \bibinfo{booktitle}{\emph{Proceedings of the AAAI Conference on Artificial Intelligence}}, Vol.~\bibinfo{volume}{39}. \bibinfo{pages}{28276--28283}.
\newblock


\bibitem[Porter and Wood(2021)]%
        {Porter2021}
\bibfield{author}{\bibinfo{person}{Ethan Porter} {and} \bibinfo{person}{Thomas~J. Wood}.} \bibinfo{year}{2021}\natexlab{}.
\newblock \showarticletitle{The global effectiveness of fact-checking: Evidence from simultaneous experiments in Argentina, Nigeria, South Africa, and the United Kingdom}.
\newblock \bibinfo{journal}{\emph{Proceedings of the National Academy of Sciences}} \bibinfo{volume}{118}, \bibinfo{number}{37} (\bibinfo{year}{2021}), \bibinfo{pages}{e2104235118}.
\newblock
\showeprint{https://www.pnas.org/doi/pdf/10.1073/pnas.2104235118}
\href{https://doi.org/10.1073/pnas.2104235118}{doi:\nolinkurl{10.1073/pnas.2104235118}}


\bibitem[Qi et~al\mbox{.}(2023a)]%
        {qi2023fakesv}
\bibfield{author}{\bibinfo{person}{Peng Qi}, \bibinfo{person}{Yuyan Bu}, \bibinfo{person}{Juan Cao}, \bibinfo{person}{Wei Ji}, \bibinfo{person}{Ruihao Shui}, \bibinfo{person}{Junbin Xiao}, \bibinfo{person}{Danding Wang}, {and} \bibinfo{person}{Tat-Seng Chua}.} \bibinfo{year}{2023}\natexlab{a}.
\newblock \showarticletitle{Fakesv: A multimodal benchmark with rich social context for fake news detection on short video platforms}. In \bibinfo{booktitle}{\emph{Proceedings of the AAAI Conference on Artificial Intelligence}}, Vol.~\bibinfo{volume}{37}. \bibinfo{pages}{14444--14452}.
\newblock


\bibitem[Qi et~al\mbox{.}(2023b)]%
        {fakesv}
\bibfield{author}{\bibinfo{person}{Peng Qi}, \bibinfo{person}{Yuyan Bu}, \bibinfo{person}{Juan Cao}, \bibinfo{person}{Wei Ji}, \bibinfo{person}{Ruihao Shui}, \bibinfo{person}{Junbin Xiao}, \bibinfo{person}{Danding Wang}, {and} \bibinfo{person}{Tat-Seng Chua}.} \bibinfo{year}{2023}\natexlab{b}.
\newblock \showarticletitle{FakeSV: A Multimodal Benchmark with Rich Social Context for Fake News Detection on Short Video Platforms}. In \bibinfo{booktitle}{\emph{Proceedings of the AAAI Conference on Artificial Intelligence}}. AAAI.
\newblock


\bibitem[Qi et~al\mbox{.}(2024)]%
        {Qi_2024_CVPR}
\bibfield{author}{\bibinfo{person}{Peng Qi}, \bibinfo{person}{Zehong Yan}, \bibinfo{person}{Wynne Hsu}, {and} \bibinfo{person}{Mong~Li Lee}.} \bibinfo{year}{2024}\natexlab{}.
\newblock \showarticletitle{SNIFFER: Multimodal Large Language Model for Explainable Out-of-Context Misinformation Detection}. In \bibinfo{booktitle}{\emph{Proceedings of the IEEE/CVF Conference on Computer Vision and Pattern Recognition (CVPR)}}. \bibinfo{pages}{13052--13062}.
\newblock


\bibitem[Ravi et~al\mbox{.}(2024)]%
        {ravi2024sam}
\bibfield{author}{\bibinfo{person}{Nikhila Ravi}, \bibinfo{person}{Valentin Gabeur}, \bibinfo{person}{Yuan-Ting Hu}, \bibinfo{person}{Ronghang Hu}, \bibinfo{person}{Chaitanya Ryali}, \bibinfo{person}{Tengyu Ma}, \bibinfo{person}{Haitham Khedr}, \bibinfo{person}{Roman R{\"a}dle}, \bibinfo{person}{Chloe Rolland}, \bibinfo{person}{Laura Gustafson}, {et~al\mbox{.}}} \bibinfo{year}{2024}\natexlab{}.
\newblock \showarticletitle{Sam 2: Segment anything in images and videos}.
\newblock \bibinfo{journal}{\emph{arXiv preprint arXiv:2408.00714}} (\bibinfo{year}{2024}).
\newblock


\bibitem[Shang et~al\mbox{.}(2025)]%
        {10854802}
\bibfield{author}{\bibinfo{person}{Lanyu Shang}, \bibinfo{person}{Yang Zhang}, \bibinfo{person}{Yawen Deng}, {and} \bibinfo{person}{Dong Wang}.} \bibinfo{year}{2025}\natexlab{}.
\newblock \showarticletitle{MultiTec: A Data-Driven Multimodal Short Video Detection Framework for Healthcare Misinformation on TikTok}.
\newblock \bibinfo{journal}{\emph{IEEE Transactions on Big Data}} (\bibinfo{year}{2025}), \bibinfo{pages}{1--18}.
\newblock
\href{https://doi.org/10.1109/TBDATA.2025.3533919}{doi:\nolinkurl{10.1109/TBDATA.2025.3533919}}


\bibitem[Shao et~al\mbox{.}(2023)]%
        {shao2023dgm4}
\bibfield{author}{\bibinfo{person}{Rui Shao}, \bibinfo{person}{Tianxing Wu}, {and} \bibinfo{person}{Ziwei Liu}.} \bibinfo{year}{2023}\natexlab{}.
\newblock \showarticletitle{Detecting and Grounding Multi-Modal Media Manipulation}. In \bibinfo{booktitle}{\emph{IEEE Conference on Computer Vision and Pattern Recognition (CVPR)}}.
\newblock


\bibitem[Shao et~al\mbox{.}(2024)]%
        {shao2024dgm4++}
\bibfield{author}{\bibinfo{person}{Rui Shao}, \bibinfo{person}{Tianxing Wu}, \bibinfo{person}{Jianlong Wu}, \bibinfo{person}{Liqiang Nie}, {and} \bibinfo{person}{Ziwei Liu}.} \bibinfo{year}{2024}\natexlab{}.
\newblock \showarticletitle{Detecting and Grounding Multi-Modal Media Manipulation and Beyond}.
\newblock \bibinfo{journal}{\emph{IEEE Transactions on Pattern Analysis and Machine Intelligence (TPAMI)}} (\bibinfo{year}{2024}).
\newblock


\bibitem[Soucek and Lokoc(2024)]%
        {soucek2024transnet}
\bibfield{author}{\bibinfo{person}{Tom{\'a}s Soucek} {and} \bibinfo{person}{Jakub Lokoc}.} \bibinfo{year}{2024}\natexlab{}.
\newblock \showarticletitle{Transnet v2: An effective deep network architecture for fast shot transition detection}. In \bibinfo{booktitle}{\emph{Proceedings of the 32nd ACM International Conference on Multimedia}}. \bibinfo{pages}{11218--11221}.
\newblock


\bibitem[Vaccari and Chadwick(2020)]%
        {vaccari2020deepfakes}
\bibfield{author}{\bibinfo{person}{Cristian Vaccari} {and} \bibinfo{person}{Andrew Chadwick}.} \bibinfo{year}{2020}\natexlab{}.
\newblock \showarticletitle{Deepfakes and disinformation:ma Exploring the impact of synthetic political video on deception, uncertainty, and trust in news}.
\newblock \bibinfo{journal}{\emph{Social media+ society}} \bibinfo{volume}{6}, \bibinfo{number}{1} (\bibinfo{year}{2020}), \bibinfo{pages}{2056305120903408}.
\newblock


\bibitem[Wang et~al\mbox{.}(2024)]%
        {wang2024official}
\bibfield{author}{\bibinfo{person}{Yihao Wang}, \bibinfo{person}{Lizhi Chen}, \bibinfo{person}{Zhong Qian}, {and} \bibinfo{person}{Peifeng Li}.} \bibinfo{year}{2024}\natexlab{}.
\newblock \showarticletitle{Official-NV: An LLM-Generated News Video Dataset for Multimodal Fake News Detection}.
\newblock \bibinfo{journal}{\emph{arXiv preprint arXiv:2407.19493}} (\bibinfo{year}{2024}).
\newblock


\bibitem[Westerlund(2019)]%
        {westerlund2019emergence}
\bibfield{author}{\bibinfo{person}{Mika Westerlund}.} \bibinfo{year}{2019}\natexlab{}.
\newblock \showarticletitle{The emergence of deepfake technology: A review}.
\newblock \bibinfo{journal}{\emph{Technology Innovation Management Review}} \bibinfo{volume}{9}, \bibinfo{number}{11} (\bibinfo{year}{2019}), \bibinfo{pages}{39--52}.
\newblock


\bibitem[Xie et~al\mbox{.}(2025)]%
        {xie-etal-2025-fire}
\bibfield{author}{\bibinfo{person}{Zhuohan Xie}, \bibinfo{person}{Rui Xing}, \bibinfo{person}{Yuxia Wang}, \bibinfo{person}{Jiahui Geng}, \bibinfo{person}{Hasan Iqbal}, \bibinfo{person}{Dhruv Sahnan}, \bibinfo{person}{Iryna Gurevych}, {and} \bibinfo{person}{Preslav Nakov}.} \bibinfo{year}{2025}\natexlab{}.
\newblock \showarticletitle{{FIRE}: Fact-checking with Iterative Retrieval and Verification}. In \bibinfo{booktitle}{\emph{Findings of the Association for Computational Linguistics: NAACL 2025}}, \bibfield{editor}{\bibinfo{person}{Luis Chiruzzo}, \bibinfo{person}{Alan Ritter}, {and} \bibinfo{person}{Lu~Wang}} (Eds.). \bibinfo{publisher}{Association for Computational Linguistics}, \bibinfo{address}{Albuquerque, New Mexico}, \bibinfo{pages}{2901--2914}.
\newblock
\showISBNx{979-8-89176-195-7}
\urldef\tempurl%
\url{https://aclanthology.org/2025.findings-naacl.158/}
\showURL{%
\tempurl}


\bibitem[Yan et~al\mbox{.}(2025)]%
        {yan2025mtparetomultimodaltargetedpareto}
\bibfield{author}{\bibinfo{person}{Kaiying Yan}, \bibinfo{person}{Moyang Liu}, \bibinfo{person}{Yukun Liu}, \bibinfo{person}{Ruibo Fu}, \bibinfo{person}{Zhengqi Wen}, \bibinfo{person}{Jianhua Tao}, \bibinfo{person}{Xuefei Liu}, {and} \bibinfo{person}{Guanjun Li}.} \bibinfo{year}{2025}\natexlab{}.
\newblock \bibinfo{title}{MTPareto: A MultiModal Targeted Pareto Framework for Fake News Detection}.
\newblock
\showeprint[arxiv]{2501.06764}~[cs.LG]
\urldef\tempurl%
\url{https://arxiv.org/abs/2501.06764}
\showURL{%
\tempurl}


\bibitem[Yao et~al\mbox{.}(2023)]%
        {yao2023end}
\bibfield{author}{\bibinfo{person}{Barry~Menglong Yao}, \bibinfo{person}{Aditya Shah}, \bibinfo{person}{Lichao Sun}, \bibinfo{person}{Jin-Hee Cho}, {and} \bibinfo{person}{Lifu Huang}.} \bibinfo{year}{2023}\natexlab{}.
\newblock \showarticletitle{End-to-end multimodal fact-checking and explanation generation: A challenging dataset and models}. In \bibinfo{booktitle}{\emph{Proceedings of the 46th International ACM SIGIR Conference on Research and Development in Information Retrieval}}. \bibinfo{pages}{2733--2743}.
\newblock


\bibitem[Zhang et~al\mbox{.}(2020)]%
        {zhang2020does}
\bibfield{author}{\bibinfo{person}{Zhu Zhang}, \bibinfo{person}{Zhou Zhao}, \bibinfo{person}{Yang Zhao}, \bibinfo{person}{Qi Wang}, \bibinfo{person}{Huasheng Liu}, {and} \bibinfo{person}{Lianli Gao}.} \bibinfo{year}{2020}\natexlab{}.
\newblock \showarticletitle{Where does it exist: Spatio-temporal video grounding for multi-form sentences}. In \bibinfo{booktitle}{\emph{Proceedings of the IEEE/CVF Conference on Computer Vision and Pattern Recognition}}. \bibinfo{pages}{10668--10677}.
\newblock


\bibitem[Zhong et~al\mbox{.}(2024)]%
        {zhong2024vmidmultimodalfusionLLM}
\bibfield{author}{\bibinfo{person}{Weihao Zhong}, \bibinfo{person}{Yinhao Xiao}, \bibinfo{person}{Minghui Xu}, {and} \bibinfo{person}{Xiuzhen Cheng}.} \bibinfo{year}{2024}\natexlab{}.
\newblock \bibinfo{title}{VMID: A Multimodal Fusion LLM Framework for Detecting and Identifying Misinformation of Short Videos}.
\newblock
\showeprint[arxiv]{2411.10032}~[cs.CV]
\urldef\tempurl%
\url{https://arxiv.org/abs/2411.10032}
\showURL{%
\tempurl}


\end{thebibliography}

\clearpage

\end{document}